\documentstyle[floats,twocolumn,prl,aps]{revtex}
\begin{document}
\draft
\title{
Free Energy Functional for
  Nonequilibrium  Systems : An Exactly Solvable Case
  }
\author{B. Derrida$^{\dag }$,  
J.   L. Lebowitz$^{\ddag*}$ and
E.   R. Speer$^{\ddag}$}

\address{$\dag$ Laboratoire de Physique Statistique,
 24 rue Lhomond, 75231 Paris Cedex 05,
 France;  \\
$\ddag$  Department of Mathematics,
Rutgers University, New Brunswick, NJ 08903; \\
$*$ Also  Department of Physics,
Rutgers University, New Brunswick, NJ 08903; \\
{\rm emails: derrida@lps.ens.fr,
  lebowitz@math.rutgers.edu, speer@math.rutgers.edu}}

\author{\parbox{397pt}{\vglue 0.3cm \small
We  consider 
the steady state of an open system in which there is a  flux of matter 
between two  reservoirs at different  chemical potentials.
For a large system of size $N$,
 the  probability of  any  macroscopic density profile $\rho(x)$  is 
$\exp[-N{\cal F}(\{\rho\})]$; 
${\cal F}$ thus generalizes to nonequilibrium systems the notion of
free energy density for equilibrium systems. 
  Our exact expression for $\cal  F$
is a nonlocal  functional of
$\rho$, which yields the macroscopically long range correlations in
the nonequilibrium steady state previously predicted by fluctuating hydrodynamics and observed experimentally.
\\
 }}
\maketitle


The extension of the central object of equilibrium
statistical mechanics, entropy or free energy, to nonequilibrium
systems in which there is a
transport of matter or energy has been the holy grail
of nonequilibrium statistical mechanics since the time of
Boltzmann. An important step in that direction was taken
by Onsager and Machlup \cite{OM} for linear deviations from
equilibrium, and there have been many further extensions to time dependent
evolutions starting and remaining in local equilibrium
\cite{KOV,KL,G,E,BDGJL}.
The extension to stationary nonequilibrium states,
for which one
has no a priori control of how close the system is 
to local equilibrium,
has however remained difficult.
One knows from approximate theories like fluctuating
hydrodynamics \cite{DKS}  that such states exhibit long range correlations
totally absent from equilibrium systems (even at the critical point).
These correlations have been measured experimentally in a fluid with a steady
heat current \cite{Exp}. Their derivation from a well defined macroscopic
functional valid beyond local equilibrium is clearly an essential
step in understanding pattern formation in 
more general stationary nonequilibrium
states such as the B\'enard
system. We report here what we believe is the first  exact 
derivation of such a functional for a nonequilibrium model which is
relatively simple but 
exhibits
the realistic feature of macroscopically long range correlations.

For an equilibrium system, such as a lattice gas, in  
a unit cube containing $L^d$  sites 
with spacing $1/L$  (or a
similar continuum system), at temperature $T$ and
chemical potential $\nu$, 
 the probability of observing
a specified macroscopic density profile, with density $\rho(x)$ at
macroscopic position $x$ in the unit cube,  is given by
 $P(\{\rho(x)\}) \sim \exp\big[-L^d{\cal F}_{\rm eq}(\{\rho\})]$.  Here
 ${\cal F}_{\rm eq}(\{\rho\})
    =\int\big[f(\rho(x))-f(\bar\rho(x))\big]\,dx$,
 where the integration is over the unit cube,
with $f(\rho)$ the grand canonical free energy density and $\bar\rho$ 
the
equilibrium density profile, obtained by minimizing
$\int f(\rho(x))\,dx$\cite{HS}.
The profile $\bar\rho(x)$ will be independent of $x$
unless there is an external potential.  We have
suppressed the dependence of $f$ and ${\cal F}_{\rm eq}$ on the constant
temperature $T$, and assume that neither $\rho$ nor $\bar\rho$ passes 
through
a phase transition region at this temperature.

In this letter we generalize the expression for ${{\cal 
F}}_{\rm eq}$ to
the case of a  system maintained, by contact with two
boundary reservoirs at unequal chemical potentials $\nu_0$ and
$\nu_1$,  in a stationary nonequilibrium state 
with a constant particle flux.  We consider
perhaps the simplest such system, the one-dimensional
symmetric simple exclusion process
on a lattice of $N$ sites with open boundaries\cite{ELS}.
Each site $i$, $i=1,\ldots,N$, is either empty ($\tau_i=0$)
or occupied by a single particle
($\tau_i=1$). 
Each particle independently
attempts to jump to its right neighboring site,
and to its left neighboring site, in each case at rate $1$.
It succeeds if the target site is empty; otherwise nothing happens.
At the  boundary sites, $1$ and $N$, particles
are added or removed: a particle is
added to site $1$, when the site is empty, at rate $\alpha$,
 and removed, when the site is occupied, at rate $\gamma$; similarly
particles are added to site $N$ at rate $\delta$ and removed at rate 
$\beta$.
We can think of sites
$i=0$ and $i=N+1$ as occupied with probabilities 
$\rho_0=\exp \nu_0 /(1+\exp \nu_0 )=\alpha/(\alpha + \gamma)$ and 
$\rho_1=\exp \nu_1 /(1+\exp \nu_1 )=\delta/(\beta + \delta)$,
independent of $\tau$. 
We will assume for
{%
definiteness that $\rho_0 > \rho_1$, but of
}%
course, due to the left-right symmetry, this is not
a restriction.

The probabilities of the microscopic configurations
in the steady state may be
calculated through the
{%
 so-called
}%
 matrix method \cite{DEHP}.
Here we ask for the probability
of seeing a specified macroscopic density profile $\rho(x)$, where
 $0\le x\le1$ and $0\le\rho(x)\le1$.
By definition, this is the sum of the probabilities of all
 microscopic configurations consistent with the given
 profile. We shall not give a precise definition 
 (see \cite{KL,HS})
of ``consistent'' here;
 roughly speaking we include all configurations $\tau$ such that 
{%
  for any $y,z$ with $0\le y<z\le1$,
 $\left|{1 \over N} \sum_{i=yN}^{zN}\tau_i-\int_y^z
\rho(x)\,dx\right|<\delta_N$, with
}%
 $\delta_N\to0$ as $N\to\infty$.

 Our   main result  is
a parametric formula for this probability: for large  $N$,
$P(\{\rho(x)\}) \sim \exp\big[-N{\cal F}(\{\rho\})]$,
with
\begin{eqnarray}
{\cal F}(\{\rho\})&\equiv& \int_0^1 dx
   \left\{ \rho(x) \log \left({ \rho(x) \over F(x)}\right) \right.
     \nonumber\\
 && \hskip-31pt
  + \left. (1- \rho(x)) \log \left( 1 - \rho(x) \over 1 -F(x) \right)
  + \log \left({ F'(x) \over \rho_1 - \rho_0} \right) \right\}.
\label{simple_expression}
\end{eqnarray}
Here $F$ is an auxiliary function
determined by  $\rho(x)$: it is the
{%
decreasing
}%
 solution of the  differential equation
\begin{equation}
\rho(x) =
 F(x) + {F(x)[1-F(x)] F''(x) \over F'(x)^2},
\label{rho(t)}
\end{equation}
satisfying  the boundary conditions
\begin{equation}
 F(0)= \rho_0 \;,\ \ \ \ \ \ \ \ \ F(1)= \rho_1 \;. \label{condition5}
\end{equation}
It can be shown that such a solution exists and is
(at least 
{%
when $\rho_1>0$ and $\rho_0<1$) unique
}%
 \cite{DLS}.

{%
We note that if one looks for a  monotone function $F$,
satisfying the constraint (\ref{condition5}), for which the expression
${\cal G}(\{\rho\},\{F\})$ given by the right hand side of
(\ref{simple_expression})
is stationary,
one obtains  (\ref{rho(t)}) as an
Euler-Lagrange
equation:
 \begin{equation}
{\delta {\cal G}(\{\rho\},\{F\}) \over \delta F(x) } = 0.
  \label{Euler-Lagrange}
 \end{equation}
   We can in fact prove that this stationary point is a maximum\cite{DLS}
when $0 < \rho_1 < \rho_0 < 1$ and
expect this to be true in general:
 \begin{equation}
{\cal F(\{\rho\})}
  = \sup_F{\cal G}(\{\rho\},\{F\}).
 \label{GF}
 \end{equation}
}%

Before explaining the main steps of the
derivation of (\ref{simple_expression}-\ref{condition5}),
let us comment some of their consequences:

 (a) By its very nature ${\cal F}$ satisfies
 ${\cal F}(\{\rho\})\ge0$, with equality only for
$\rho(x)=\bar\rho(x)$,
where $\bar\rho(x)=\rho_0(1-x)+\rho_1x$
 is the profile obtained with probability one in the limit $N\to\infty$.
Any other profile  
will have ${\cal F}(\{\rho\}) > 0$ and thus, for large $N$,
exponentially small probability.  
{%
When $\rho_0 = 1$ or $\rho_1 = 0$ there
}%
are some profiles for which ${\cal F} = +\infty$; their 
probability is super-exponentially small in $N$.  For examples, see (b)
below and \cite{DLS}.

(b) For a constant profile $\rho(x)=r$, 
$F$ satisfies $F'= A F^r(1-F)^{1-r}$, where  $A$ is
fixed by (\ref{condition5}), and
\begin{equation}
{\cal F}(\{\rho\})  = \log \left[  \int_{\rho_0}^{\rho_1} \left( r 
\over z \right)^r \  \left( 1-r \over 1 - z \right)^{1-r} \ {dz \over 
\rho_1 - \rho_0} \right]\;.
\label{constantrho}
\end{equation}
We see that ${\cal F}(\{\rho\})= \infty$
if $r=1$ 
{%
and $\rho_1=0$, or $r=0$ and $\rho_0=1$.

 (c) Using (\ref{Euler-Lagrange}), one finds immediately that
 \begin{equation}
{\delta {\cal F} \over \delta \rho(x)}
  = \log\left[{\rho(x) \over F(x)}
  \cdot {1- F(x) \over 1- \rho(x)}\right].\label{vd}
 \end{equation}
This expression
}%
 can then be used to find 
optimal profiles subject to various constraints.
For example, setting (\ref{vd}) to zero implies that
$\rho(x)=F(x)$, leading to the most likely profile $\bar\rho(x)$ given 
in (a).

 (d) If we minimize ${\cal F}$ subject to the constraint of a fixed mean
density $\int_0^1 \rho(x)dx$,  the right hand side of (\ref{vd})
becomes an arbitrary constant, and
together with (\ref{rho(t)}) one obtains   that the most likely profile 
is
exponential: $\rho(x)=A_1\exp(\theta x)+ A_2$,
 the constants being 
determined by the
value of the mean density and the boundary conditions 
(\ref{condition5}).  
   (This exponential form, which is the stationary
solution of a diffusion equation with drift, was first suggested to us, 
for
some special cases, by Errico Presutti).

 Similarly, if
{%
we impose a fixed mean density in $k$ nonoverlapping intervals,
with no other 
constraints,
the optimal
profile  is exponential  inside these
intervals,  linear outside, and  in general not
continuous at the end points of the intervals.
}%

(e) When the chemical potentials of the two reservoirs are equal,
i.e.,  $\rho_0=\rho_1$, the system is in true equilibrium with
$\bar\rho(x)=\rho_1$. Eqs. (\ref{simple_expression}--\ref{condition5}) have
a well-defined 
{%
limit for $\rho_1\nearrow\rho_0$, with 
}%
$F(x)=\rho_0 + (\rho_1 - \rho_0) x + O( (\rho_1 - \rho_0)^2) $.
It is also natural to consider a local 
equilibrium Gibbs measure corresponding to
a spatially varying chemical
potential\cite{KOV,HS} which is adjusted to
maintain the same optimal profile $\bar\rho(x)$.
 For this system the 
large
deviation functional (free energy) is just ${\cal F}_{\rm 
eq}(\{\rho\})$,
which has the explicit form
 \begin{eqnarray}
 {\cal F}_{\rm eq}(\{\rho\})&=&
   \int\bigg\{ \rho(x) \log {\rho(x) \over \bar \rho(x)}\nonumber\\
  &&\hskip20pt +\;[1- \rho(x)]\log{(1 - \rho(x))
         \over(1 - \bar \rho(x))}\bigg\}\,dx.
   \label{equil}
 \end{eqnarray}
{%
From (\ref{GF}):
${\cal F(\{\rho\})}\ge{\cal G}(\{\rho\},\{\bar\rho\})
  ={\cal F}_{\rm eq}(\{\rho\})$, so that
   expressions (\ref{simple_expression})
and (\ref{equil}) are in general different
 \begin{equation}
 {\cal F}(\{\rho\})\ge{\cal F}_{\rm eq}(\{\rho\}),\label{FFeq}
 \end{equation}
 with equality only for
$\rho(x)=\bar\rho(x)$ or $\rho_0=\rho_1$;
}%

(f)
Using the fact that the exponential is the  optimal profile
for the case of a fixed mean density in the
entire interval, we may compute the distribution of
$M$, the total number of particles in the system, in the steady state 
for
large $N$.  We find that  the fluctuations of $M$  predicted by
(\ref{simple_expression}) are reduced in
comparison to those in a system in local equilibrium
(\ref{equil})  with the same $\bar\rho$:
 \begin{eqnarray}
  \lim_{N\to\infty}N^{-1}
    \bigl[\langle M^2\rangle_{\rm SNS}-\langle 
M\rangle^2\bigr]\nonumber\\
 &&\hskip-115pt =\;
   \lim_{N\to\infty}N^{-1}
    \bigl[\langle M^2\rangle_{\rm  eq}-\langle M\rangle^2\bigr]
      - {(\rho_1 - \rho_0)^2 \over 12} .
\label{variance}
\end{eqnarray}
 We may also obtain (\ref{variance}) by expanding $\rho(x)$ about
$\bar\rho(x)$ in (\ref{simple_expression}).  The result agrees with that
obtained in \cite{HS1} directly from the microscopic model and from
fluctuating hydrodynamics\cite{DKS}.

{\it Derivation:}
let us now
 sketch the derivation  of (\ref{simple_expression}-\ref{condition5}).
The probability of a configuration $\tau =
\{\tau_1,\dots,\tau_N\}$ in
the steady state of our model is given by \cite{DEHP}
 \begin{equation}
   P_N(\tau) = {\omega_N(\tau)
    \over  \langle W|(D+E)^N|V\rangle}\;, \label{prob}
 \end{equation}
where the weights $\omega_N(\tau)$ are given by
 \begin{equation}
   \omega_N(\tau) = \langle W|\Pi_{i=1}^N (\tau_i D + (1 - 
\tau_i)E)|V\rangle 
    \label{omega}
 \end{equation}
  and the matrices $D$ and $E$ and the vectors $|V\rangle$ and $\langle 
W|$ satisfy
 \begin{eqnarray}
     DE - ED &=& D + E  \label{DE}\;,\\
  (\beta D - \delta E)|V\rangle = |V\rangle\;,\ &&
   \langle W|(\alpha E - \gamma D) = \langle W|\;. \label{alg}
 \end{eqnarray}
{%
 Although  proving that (\ref{prob}-\ref{alg}) do give the 
weights in the steady state is rather easy \cite{DEHP}, there is not so far
 a simple physical interpretation of the matrices 
$D$ and $E$ or of the vectors $|V \rangle$ or $\langle W|$.

 To obtain the probability  
 $P_{N_1,\dots,N_n}(M_1,M_2,\dots,M_n)$ 
that
   $M_1$
particles  are located on the first $N_1$ sites, $M_2$ particles
on the next $N_2$ sites, etc.., 
 we first 
calculate the sum  $\Omega_{N_1,\dots,N_n}(M_1,M_2,\dots,M_n)$ of the 
weights of all the corresponding configurations.
}%
The key to obtaining  (\ref{simple_expression})  is that
 the following generating function,
which plays the role of the grand-canonical-pressure partition function,
can be computed exactly:
\begin{eqnarray}
 Z(\lambda_1,\ldots,\lambda_n; \mu_1,\ldots, \mu_n) &&\nonumber \\
 &&\hskip-98pt\equiv
  \sum {\mu_1 ^{N_1}\over N_1!}\cdots{\mu_n^{N_n}\over N_n!}
 \lambda_1 ^{M_1} \cdots \lambda_n ^{M_n} 
{\Omega_{N_1,\dots,N_n}(M_1,\ldots,M_n) \over \langle W | V \rangle}
   \nonumber\\
 &&\hskip-98pt =
 { \langle W | e^{\mu_1 \lambda_1 D + \mu_1 E}  \cdots e^{\mu_n 
\lambda_n D + \mu_n E} | V \rangle  \over
\langle W |  V \rangle }\;,
\label{Zdef}
 \end{eqnarray}
where  the sum is over all $N_i,M_i\ge0$
{%
and the parameters $\mu_i$ and $\lambda_i$ are
 conjugate to the $N_i$ and $M_i$.
}%
To  do the calculation we used repeatedly the following identity, which 
follows from (\ref{DE}),
\begin{equation}
 e^{x D + y E} = \left( (x-y) e^y \over x e^y - y e^x \right)^E
\   \left( (x-y) e^x \over x e^y - y e^x \right)^D\;,
\label{identity1}
\end{equation}
 and the result is that
 \begin{eqnarray}
 Z = \left(\rho_0 - \rho_1 \over g \right)^{a+b} \exp \left[ a 
\sum_{i=1}^n \mu_i(1-\lambda_i) \right] \;,
\label{Zresult}
 \end{eqnarray}
where  $a = (\alpha + \gamma)^{-1}$, $b =
(\beta + \delta)^{-1}$, and
 \begin{eqnarray}
 g &=& \biggl[-\rho_1 + \rho_0 e^{\sum_{i=1}^n \mu_i(1-\lambda_i)} 
\nonumber\\
 &&\hskip5pt+\sum_{i=1}^n {1 \over \lambda_i - 
1}(e^{\mu_i(1-\lambda_i)} - 1)
e^{\sum_{j>i} \mu_j(1-\lambda_j)}\biggr],
\label{gdef}
 \end{eqnarray}
All the rest of the derivation consists in extracting from the exact 
expression
(\ref{Zresult})  the
behavior of $ \Omega$ and $P$ for large $N_i$'s.

{%
First  (\ref{Zresult}) gives the normalization factor in 
(\ref{prob}),
\begin{equation}
 { \langle W | (D+E)^N | V \rangle  \over
\langle W |  V \rangle } =  {\Gamma(a+b+N) \over \Gamma(a+b) (\rho_0 - 
\rho_1)
^N}\;,
\label{Z0}
\end{equation}
so that  $P_{N_1,\dots,N_n}(M_1,M_2,\dots,M_n)$  is given simply by
 $\Omega_{N_1,\dots,N_n}(M_1,M_2,\dots,M_n)$  divided by (\ref{Z0}).

Now the  large $N_i$ behavior of $\Omega_{N_1,\ldots,N_n}$
controls the position and the nature of the singularities of
$Z$ closest to the origin $\mu_i=\lambda_i=0$; conversely, this relation
can be inverted \cite{DLS} to determine the asymptotic behavior of 
$\Omega_{N_1,\ldots,N_n}$ from the equation 
$g=0$ of the surface on which $Z$ is singular, just as the growth of the
coefficients of a power series of one variable is determined by the
singularity nearest the origin. 
}%
In particular,
if one sets $N_j= N y_j$  and $M_j = N_j r_j$ then, for large $N$
and fixed $y_i$, $r_i$,
 \begin{eqnarray}
 {\log P_{N_1,\dots,N_n}(M_1,\dots,M_n) \over N} \nonumber\\
 &&\hskip-100pt\simeq \;
   \log(\rho_0 - \rho_1)
   - \sum_{j=1}^n y_j\big(\log {\mu_j/ y_j} + r_j \log \lambda_j
\big)\;,\label{f}
 \end{eqnarray}
where $y_j$ and $r_j$ are
related to the parameters $\mu_1,\dots,\mu_n$ and
$\lambda_1,\dots,\lambda_n$ by
 \begin{equation}
 y_j = {{\partial g \over \partial \log \mu_j} \over
   \sum_{i=1}^n {\partial    g \over \partial \log \mu_i}}\;,
  \qquad
   r_j = {{\partial g \over \partial \log \lambda_j} \over
   { \partial g \over\partial \log \mu_j}}\;,\label{yjrj}
\end{equation}
 with all derivatives calculated on the manifold $g = 0$.

Equations (\ref{f}) and (\ref{yjrj})
determine $\cal F$ in a parametric form.  As the
$\mu_j$'s and the $\lambda_j$'s vary, they give the sizes of the boxes
$y_j$, their particle densities $r_j$, and the corresponding 
probabilities.
Note that the parameters $a$ and $b$ do not appear
{%
in (\ref{gdef})
and therefore in the equation $g=0$;
so for large $N$,
}%
 only
$\rho_0$ and $\rho_1$ remain relevant.

This parametric form can be simplified by replacing the role of the
$\mu_i$ and $\lambda_i$ by a single sequence of
parameters $G_i$.  Let us define the constant $C$ by
$C = \sum_{i=1}^n \partial g/ \partial \log \mu_i$
and the sequence $G_i$  by
 \begin{equation}
   G_i =  C^{-1}e^{\sum_{j=i}^n \mu_j(1-\lambda_j)}\;, \qquad 
G_{n+1} = C^{-1}\;. \label{Gi}
 \end{equation}
 It follows from (\ref{Gi}) that
 $\mu_i = \log(G_i/G_{i+1})/(1 - \lambda_i)$
and
 \begin{eqnarray}
 {1 \over \lambda_i - 1}
  &=& {1 \over G_{i+1}}{y_i \over \log(G_i/G_{i+1})} - \rho_0 
\nonumber\\
 &&\hskip20pt + \sum_{j=1}^i \big ({1 \over G_j} - {1 \over G_{j+1}} 
\big)
   {y_j   \over \log(G_j/G_{j+1})}\;.
\label{li-1}
 \end{eqnarray}
The condition that $g=0$ becomes
 \begin{equation}
 \rho_0 - \rho_1
   = \sum_{j=1}^n \big({1 \over G_j} - {1 \over G_{j+1}} \big ){y_j
       \over \log(G_j/G_{j+1})}\;,
\label{norma}
 \end{equation}
which can be thought of as an equation which determines $G_{n+1}$ in 
terms of
$G_1,\ldots,G_n$.

Once the $\lambda_i$'s are known, one gets for the
$r_i$'s and the large deviation function
 \begin{equation}
 r_i = -{\lambda_i \over 1 - \lambda_i} - {\lambda_i \over (1 -
\lambda_i)^2}{G_i - G_{i+1} \over y_i} \label{r}
 \end{equation}
and
 \begin{eqnarray}
{\log[P_{N_1,\dots,N_n}(M_1,\dots,M_n)]\over N} &&\nonumber\\
 &&\hskip-50pt =\, -\sum_{i=1}^n \Big \{y_i \log \Big [{\log({G_i 
\over G_{i+l}}) \over
y_i(\rho_0 - \rho_1)} \Big ] \label{P1}\\
 &&\hskip-30pt - y_i \log(1 - \lambda_i) + y_i r_i \log(\lambda_i) \Big\}.
 \nonumber
 \end{eqnarray}

To take the limit $N \to \infty$ followed by
$n \to \infty$, $y_i \to 0$, we
introduce a continuous variable $x$, $0 \leq x \leq 1$, and let
$x_i = y_1 + y_2 + \dots + y_i$.
All the discrete sequences can now be thought of as functions of
$x_1,\ldots,x_n$ with
$G_i \equiv G(x_i)$, $\lambda_i\equiv\lambda(x_i)$ and
$r_i \equiv \rho(x_i)$, $i=1,\ldots,n$.
Then, extending $G$, $\lambda$, and $\rho$ to be smooth functions
of $x$, so that $G_{i+1} - G_i \simeq y_i G^\prime(x_i)$,
one finds that (\ref{li-1},\ref{norma}) become
 \begin{eqnarray}
  {1 \over \lambda(x) - 1} &=& {-1 \over G^\prime(x)} - \rho_0 - 
  \int_0^x {du \over G(u)}\;,\label{lambda}\\
  \rho_1 &=& \rho_0 + \int_0^1 {du \over G(u)}\;. \label{Gbd}
 \end{eqnarray}
 At this stage it is more convenient to replace
$G(x)$ by the  function $F(x)$ defined by
$F(x) = \rho_0 + \int_0^x G(u)^{-1}\,du$.
The expression (\ref{lambda})  for $\lambda(x)$ becomes then
 \begin{equation}
 {1 \over \lambda(x) - 1} = {F'(x)^2 \over F^{\prime \prime}(x)} - 
F(x).
 \label{lambdaF}
 \end{equation}
 Using the above relations
we may rewrite (\ref{r}) and (\ref{P1}) in terms of $F$,
obtaining
(\ref{simple_expression}) and (\ref{rho(t)}); 
the boundary conditions (\ref{condition5}) come from
(\ref{Gbd}).  The monotonicity of $F$ follows from the
uniqueness of the sign of
$G$ (see  (\ref{Gi})).

{\it Possible extensions:}
{%
It would be nice  to give a 
physical interpretation of the 
 Euler-Lagrange equations (\ref{Euler-Lagrange},\ref{GF})
 directly 
 from a  macroscopic or semimacroscopic approach.
}%
Such an approach  might
allow an extension to the large deviation function for
time-dependent profiles and permit one to unify various results recently 
obtained for the large deviation functions of time dependent profiles 
or currents \cite{BDGJL,DL,PS}.

The matrix method
(with modified rules (\ref{DE}-\ref{alg})) 
also applies
\cite{DEHP}
to  the asymmetric simple
exclusion process, in which particles jump to the left and right at
different rates. 
{%
We are at present trying 
  to generalize our  
calculations to this case.
We  have  already
calculated the probability of
a
specified mean density $\rho$ (total number of particles) 
for the special case in
which particles jump only to their right and
with only input, with rate $\alpha>1$, at the left and output, with
rate
$\beta>1 $, at the right:
}%
 \begin{equation}
 N^{-1} \log P(\rho)  \sim
  -2[ \rho \log 2 \rho +  (1- \rho) \log 2 (1- \rho) ]\;.
 \end{equation}
 For a Bernoulli measure at uniform density $1/2$, the result
would
be the same except for the factor $2$ in front of the whole expression,
which
here again expresses the fact that fluctuations are reduced by long range
correlations.

{\it Conclusion:}
{%
For the simple model we studied here,   
the large deviation function
}%
 ${{\cal F}}$  (which extends the notion of  free energy to 
nonequilibrium systems)
 is a nonlocal functional 
 (\ref{simple_expression}--\ref{condition5}) of the density $\rho$.
 This  implies that the probability of a given profile is not
the product of probabilities for, say, the left and right halves of the
system,  in contrast to the situation for 
equilibrium systems.
Since long range correlations are
 expected for general stationary nonequilibrium states
 and  have even been measured experimentally \cite{Exp}, 
the nonlocal nature of 
${{\cal F}}(\{\rho\})$ is presumably also  general, 
in contrast with the conjecture
\cite{G,E}
that 
$\cal F$ is a local function of the 
hydrodynamic variables.

 We thank E. Presutti for very helpful discussions.  The work of
J.~L.~Lebowitz was supported by NSF Grant DMR--9813268, AFOSR Grant
F49620/0154, DIMACS and its supporting agencies, and NATO Grant
PST.CLG.976552.  J.L.L. and E.~R.~Speer acknowledge the hospitality of the
I.H.E.S. in the spring of 2000, where this work was begun.

 
 \end{document}